\documentclass[twocolumn,prb,amsmath,amssymb,showpacs,
superscriptaddress,floatfix]{revtex4}

\usepackage{graphicx}

\usepackage{hyperref}

\DeclareMathOperator{\sgn}{sign}

\renewcommand{\Im}{\mathop\mathrm{Im}\nolimits}
\newcommand{\si}{\mathop{\rm si}\nolimits}
\newcommand{\ci}{\mathop{\rm ci}\nolimits}

\begin{document}

\title{Domain wall effects in ferromagnet-superconductor structures.}

\author{Igor~S. Burmistrov}
\affiliation{L.D.\ Landau Institute for Theoretical Physics, Russian Academy of Sciences,
117940 Moscow, Russia} \affiliation{Institute for Theoretical Physics, University of
Amsterdam, Valckenierstraat 65, 1018XE Amsterdam, The Netherlands }

\author{Nikolai~M. Chtchelkatchev}
\email{nms@landau.ac.ru} \affiliation{L.D.\ Landau Institute for Theoretical Physics, Russian
Academy of Sciences, 117940 Moscow, Russia} \affiliation{Institute for High Pressure Physics,
Russian Academy of Sciences, Troitsk 142092, Moscow Region, Russia}\affiliation{Moscow
Institute of Physics and Technology,  Moscow 141700, Russia}

\begin{abstract}
We investigate how domain structure of the ferromagnet in superconductor-ferromagnet
heterostructures may change their transport properties.  We calculate the distribution of
current in the superconductor induced by magnetic field of Bloch domain walls, find the
``lower critical'' magnetization of the ferromagnet that provides vortices in the
superconductor.
\end{abstract}


\maketitle

Superconductivity and ferromagnetism are two competing phenomena: while the first prefers
antiparallel spin orientation of electrons in Cooper pairs, the second forces the spins to be
aligned in parallel. Their coexistence in one and same material or their interaction in
spatially separated materials
leads to a number of new interesting phenomena, for example,
$\pi$-state of superconductor (S) -- ferromagnet (F) --
superconductor (SFS) Josephson
junctions,~\cite{Ryazanov_new,Bulaevskii,Ryazanov,Kontos,Chtch_pi,Barash,Radovic,Beenakker}
highly nonmonotonic dependence of the critical temperature $T_c$
of a SF bilayer as a function of the ferromagnet
thickness~\cite{Fominov_Ch_Golubov} and so on. Recent
investigations of SF bilayers showed that their transport
properties often strongly depend on the interplay between magnetic
structure of the ferromagnet and
superconductivity.~\cite{Bulaevskii_Chudnovsky,Sonin,Deutscher,Kinsey,Kadigrobov,Pokrovsky,Chtchelkatchev-Burmistrov,Melin,Ryazanov_vortices}
In particular, it was argued that due to ferromagnetic domains
vortices may appear in the superconducting film of the SF bilayer
and the domain configuration, in turn, may depend on the
vortices.~\cite{Pokrovsky} Recently, generation of vortices by
magnetic texture of the ferromagnet in SF junctions was
demonstrated experimentally.~\cite{Ryazanov_vortices} In a number
of experiments dealing with $T_c$ of SF bilayers, or Josephson
effect in SFS structures~\cite{Ryazanov_new} the domain
magnetizations were parallel to the SF interface. Ferromagnets
used in the experiments were often dilute with the exchange field
comparable to the superconducting gap and small domain size
[smaller or comparable to the bulk superconductor screening
length] and broad domain walls.~\cite{Ryazanov_new}

In this paper we try to put a step forward the answer to the
question, how domain structure of the ferromagnet in SF
heterostructures may change their transport properties. In the
first part of the paper we discuss the junctions where S and F are
weakly coupled (there is insulator layer in-between such that
there is no proximity effect) and magnetizations of the domains
are parallel to the SF interface. We find the distribution of
current in the superconductor induced by magnetic field of the
domain walls and the ``lower critical'' magnetization of the
ferromagnet that provides vortices in the superconductor. In the
end of the paper we estimate the critical temperature in strongly
coupled SF bilayer when the proximity effect is strong. In this
paper we do not consider the rearrangement of the domain
configuration due to the superconductor~\cite{Pokrovsky} though we
mention the superconductor induced transitions between Bloch and
Neel domain wall types. The point is that the crystal structure of
the ferromagnets in experiments~\cite{Ryazanov_new} was not
perfect. Experimental data suggests that defects, dislocations in
the lattice that appear during lithography process stick domain
configuration.

\begin{figure}[t]
\begin{center}
\includegraphics[width=75mm]{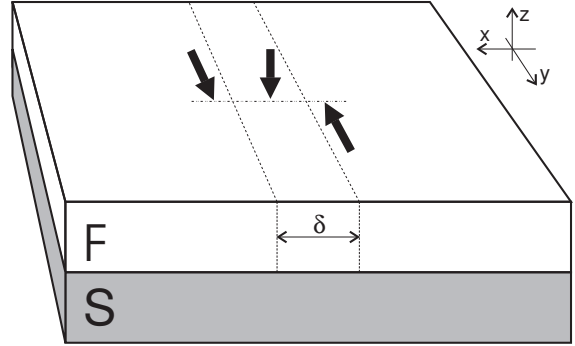}
\caption{\label{fig1} The SF junction. A sketch of a Bloch domain
wall. The magnetization rotates according to Eq.~\eqref{m}.
Magnetization in the center of the domain wall is perpendicular to
the S film. }\end{center}
\end{figure}

The domain texture of the F film is described by the following
magnetization (see Fig.~\ref{fig1})
\begin{equation}
\mathbf{M} = M \theta\left(z\right)\theta\left(d_F-
z\right)\sum\limits_{j=-\infty}^{\infty} (-1)^j \mathbf{m}(x-j L).
\end{equation}
where vector $\mathbf{m}(x)$ rotates as follows~\cite{Landau8}
\begin{equation}\label{m}
m_x=0,\quad m_y=\tanh(x/\delta),\quad m_z=-1/\cosh(x/\delta).
\end{equation}
The vector potential $\mathbf{A}$ satisfies the Maxwell-Londons equation
\begin{equation}
\nabla\times (\nabla\times\mathbf{A}) + \mathbf{A}\lambda_L^{-2}
\theta\left(-z\right)\theta\left(z+d_S\right)= 4\pi \nabla \times
\mathbf{M}\label{MLEq}
\end{equation}
that should be supplemented by the standard boundary conditions of
continuity of $\mathbf{A}$ and $\partial \mathbf{A}/\partial
z$.~\cite{Landau8} By solving Eq.~\eqref{MLEq} with a help of the
Fourier transformation we can find the distribution of the
magnetic field $\mathbf{B}=\nabla\times\mathbf{A}$ in the entire
space.~\cite{BC} In agreement with general expectations the $z$
component of the magnetization in the F film that collects at
domain walls results in the current flow in the S film. It is
convenient to define the current averaged over the thickness of
the S film, $J_y(x) = -c/(4\pi \lambda_L^2) \int_{-d_S}^{0} A_y d
z$. Then we obtain
\begin{widetext}
\begin{equation}\label{Jy}
J_y(x) =- 2\pi c M \frac{\delta}{\lambda_L^2 L}
\sum\limits_{n=0}^\infty \frac{\sin q_n x}{Q_n\cosh
(\frac{\pi}{2}q_n\delta)} \frac{Q_n \sinh Q_n d_S + q_n (\cosh
Q_nd_S-1)}{(Q_n^2 +q^2_n)\sinh Q_nd_S + 2 Q_n q_n \cosh
Q_nd_S}\left [1-\exp (-q_n d_F)\right ],
\end{equation}
\end{widetext}
where $c$ stands for the speed of light, $q_n = \pi(2n+1)/L$ and
$Q_n=\sqrt{q_n^2+\lambda_L^{-2}}$. Eq.~\eqref{Jy} constitutes one
of the principal results of the present paper. It allows to
compute the distribution of the current flow in the S film for a
general set of parameters $d_S$, $d_F$, $\delta$, $L$ and
$\lambda_L$. Below we shall analyze two the most interesting cases
of thick ($d_S, d_F\gg \lambda_L$) and thin ($d_S, d_F \ll
\lambda_L$) SF bilayer.

\emph{Thick SF bilayer.} Eq.~\eqref{Jy} can be drastically
simplified provided $d_S, d_F\gg\lambda_L$ such that the current
$J_y(x)$ becomes independent on the widths $d_S$ and $d_F$ of the
S and F films. It is given as
\begin{equation}\label{JyTk}
J_y(x) = - c M\frac{2\pi\delta}{\lambda_L^2L} \sum\limits_{n=0}^\infty\frac{\sin q_n x}{\cosh
\frac{\pi}{2}q_n\delta}
        \frac{1}{Q_n(Q_n+q_n)}.
\end{equation}

\begin{figure}[t]
\includegraphics[width=80mm]{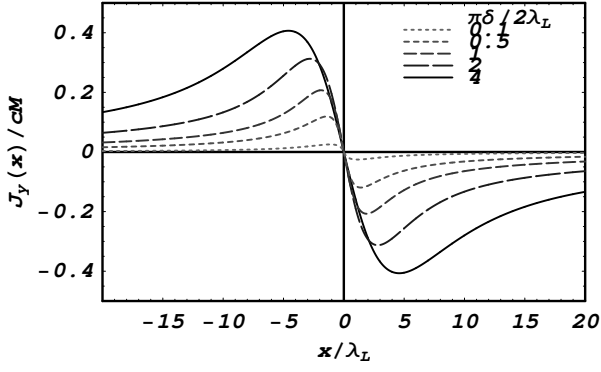}
\caption{The case $d\gg \lambda_L$. The plot of $J_y(x)/(cM)$ as a
function of $x/\lambda_L$ for
$\pi\delta/(2\lambda_L)=0.1,0.5,1,2,4$.} \label{FIGDWSThick0}
\end{figure}

In order to understand the distribution of the current $J_y(x)$ as
determined by Eq.~\eqref{JyTk} we shall first analyze the case of
a \textit{single} domain wall. Taking the limit $L\to\infty$ in
Eq.~\eqref{JyTk}, we obtain the following result for the current
in the presence of a single domain wall in the F film
\begin{equation}\label{JySW1}
\frac{J_y(x)}{cM}= -\frac{\delta}{\lambda_L}\int_0^\infty \frac{d\omega}{\sqrt{1+\omega^2}}
\frac{\sin \frac{x}{\lambda_L}\omega}{\cosh \frac{\pi\omega\delta}{2\lambda_L}}
\frac{1}{\omega+\sqrt{1+\omega^2}}.
\end{equation}
The distribution of the current $J_y(x)$ is governed by the single
parameter $\pi\delta/(2\lambda_L)$ as it is shown in
Fig.~\ref{FIGDWSThick0}. If the width $\delta$ of the domain wall
is small compared to the Londons penetration length $\lambda_L$,
$\pi\delta/(2\lambda_L)\ll 1$, we find the distribution of the
current as
\begin{equation}\label{JyTk1}
J_y(x) = c M
  \begin{cases}
    \frac{\delta}{x} \left [
\frac{|x|}{\lambda_L}K_1\left
(\frac{|x|}{\lambda_L}\right)-1\right ], & |x|\gg \delta, \\
\frac{x\delta}{2\lambda_L^2}\ln\frac{\pi\delta}{2\lambda_L},& |x|\ll\delta,
  \end{cases}
\end{equation}
where $K_1(x)$ is the modified Bessel function of the second kind. In the opposite case of the
thick domain wall, $\pi\delta/(2\lambda_L)\gg 1$, we obtain
\begin{equation}\label{JyTk2}
\frac{J_y(x)}{c M} = \tanh\frac{x}{\delta} -\frac{2}{\pi} \Im
\psi\left (\frac{1}{4} + i \frac{x}{2\pi\delta}
\right)+\frac{\lambda_L}{\delta}\frac{\tanh
\frac{x}{\delta}}{\cosh\frac{x}{\delta}},
\end{equation}
where $\psi(x)$ denotes the digamma function.

According to Eqs.~\eqref{JyTk1} and \eqref{JyTk2} the current
$J_y(x)$ behaves linearly with $x$ for $x\ll \delta$ and decays as
a power law for large $x$. The current distribution $J_y(x)$ is
spread on the length $L_s\propto \max\{\delta,\lambda_L\}$ from
the origin whereas its maximal value $J_y^m \propto cM
\delta/L_s=cM\min\{1,\delta/\lambda_L\}$.

\begin{figure}[t]
\includegraphics[width=80mm]{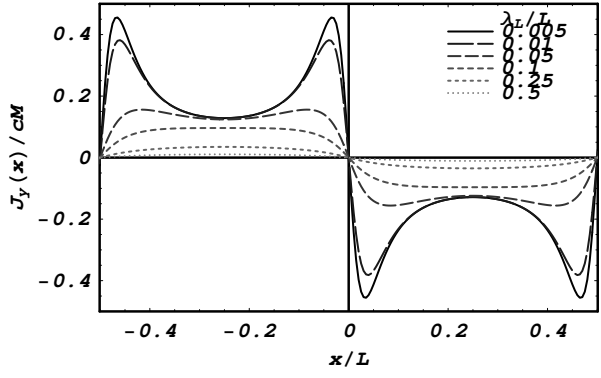}
\caption{The case $d\gg \lambda_L$. The plot of $J_y(x)/(cM)$ as a
function of $x/L$ for $\delta/L=0.02$ and
$\lambda_L/L=0.005,0.01,0.05,0.1,0.25,0.5$.} \label{FIGDWSThick1}
\end{figure}

Now we turn back to the general case of \textit{multi} domain wall
structure in the F film that corresponds to a finite size $L$ of
domains. We have evaluated the sum in Eq.~\eqref{JyTk} numerically
and present results for the current distribution in
Fig.~\ref{FIGDWSThick1}. While $\lambda_L$ remains small compared
with $L$ the profile of $J_y(x)$ corresponds to almost independent
current distributions near each domain wall that results in
distinctive two maximum structure as shown in
Fig.~\ref{FIGDWSThick1}. When $\lambda_L$ becomes of the order of
$L$ the two maximum structure transforms into sinusoidal profile
with the maximum exactly in the middle of a domain.

\begin{figure}[h]
\begin{center}
\includegraphics[width=80mm]{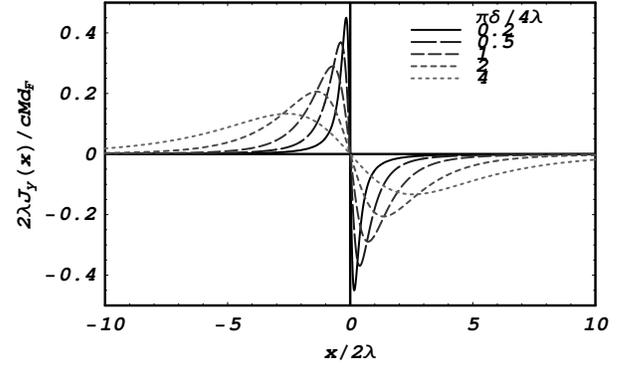}
 \caption{The case $d\ll \lambda_L$.
The plot of $2\lambda J_y(x)/(cM d_F)$ as a function of
$x/(2\lambda)$ for $\pi\delta/(4\lambda)=0.2,0.5,1,2,4$.}
\label{FIGDWSThin0}
\end{center}
\end{figure}

\emph{Thin SF bilayer.} In the case of the thin SF bilayer, $d_S,
d_F \ll \lambda_L$, by expanding the general expression~\eqref{Jy}
in powers of $d_S$ and $d_F$, we find
\begin{equation}\label{JyTn}
J_y(x)= - c M d_F\frac{2\pi \delta}{L}
\sum\limits_{n=0}^\infty\frac{\sin q_n x}{\cosh
\frac{\pi}{2}q_n\delta}
    \frac{q_n}{1+2q_n\lambda}.
\end{equation}
Here $\lambda=\lambda_L^2/d_S$ usually referred as the effective
penetration length.~\cite{Pearl} We shall first analyze the case
of a \textit{single} domain wall again. In the limit $L\to\infty$
we obtain from Eq.~\eqref{JyTn}
\begin{equation}\label{JySW1}
J_y(x) = -c M \frac{d_F \delta}{4\lambda^2}\int_0^\infty
d\omega\frac{\omega}{1+\omega}\frac{\sin
\frac{x}{2\lambda}\omega}{\cosh \frac{\pi\delta}{4\lambda}
\omega}.
\end{equation}
The distribution of the current $J_y(x)$ is presented in
Fig.~\ref{FIGDWSThin0} for different values of the parameter
$\pi\delta/(4\lambda)$.
\begin{figure}[h]
\includegraphics[width=80mm]{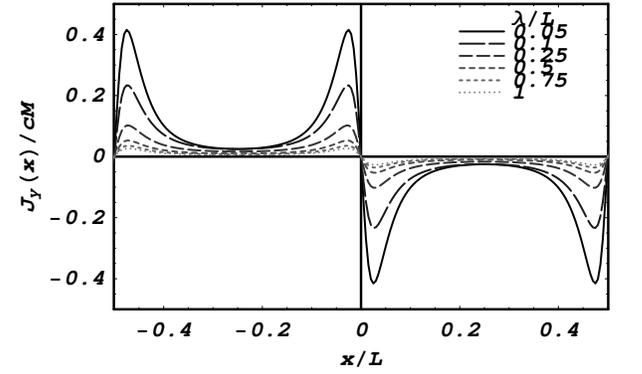}
\caption{The case $d\ll \lambda_L$. The plot of $J_y(x)/(cM)$ as a
function of $x/L$ for $\delta/L=0.02$, $d_F/L=0.1$ and
$\lambda/L=0.05,0.1,0.25,0.5,0.75,1$.} \label{FIGDWSThin1}
\end{figure}
If the domain wall is thin, $\pi\delta/(4\lambda)\ll 1$,
Eq.~\eqref{JySW1} yields
\begin{equation}
\frac{J_y(x)}{cM} =\frac{d_F}{2 \lambda} \Bigl
[\tanh\frac{x}{\delta} -\frac{2}{\pi} \Im \psi\left (\frac{1}{4} +
i \frac{x}{2\pi\delta} \right) +\frac{\delta}{2\lambda} g(x)\Bigr
],\label{JyTn1}
\end{equation}
where
\begin{equation}
g(x)= \begin{cases}
  \frac{\pi}{2}\sgn x \cos \frac{x}{2\lambda}+
 f(\frac{x}{2\lambda})
 ,& |x|\gg\delta, \\
     \frac{x}{\delta}, & |x|\ll\delta.
  \end{cases}
\end{equation}
Here $f(x)=\sin x \ci(x)-\cos x \si(x)$ with $\ci(x)$ and $\si(x)$
being the cosine and sine integral functions. In the opposite case
$\pi\delta/(4\lambda)\gg 1$ the current distribution $J_y(x)$ is
given as
\begin{equation}\label{JyTn2}
\frac{J_y(x)}{cM} = -\frac{d_F}{\delta}\frac{\tanh x/\delta}{\cosh
x/\delta}.
\end{equation}

Eqs.~\eqref{JyTn1} and \eqref{JyTn2} proves that the $J_y(x)$
increases linearly with $x$ for $x\ll \delta$ and decreases
algebraically for large $x$. The position of the maximum of
$J_y(x)$ is situated at $L_s\propto \delta$ and the value at the
maximum $J_y^m\propto (d_F/\delta)\min\{1,\delta/\lambda\}$. As
one can see therefore the current distribution for the thin SF
bilayer is qualitatively different from one for the thick SF
bilayer.

In the case of \textit{multi} domain wall structure in the F layer
we have performed evaluation of the sum in Eq.~\eqref{JyTn}
numerically and have obtained the results for the current
distribution $J_y(x)$ presented in Fig.~\ref{FIGDWSThin1}. We
mention that the two maximum structure survives even for $\lambda$
of the order of $L$ for the thin SF bilayer.

As known the lower critical field for a thin S film is much
smaller than for the bulk superconductor. Therefore it is possible
that even small magnetization collects at domain walls can induce
a vortex in the thin S film.~\cite{Pokrovsky} Let us assume that
there is a single vortex in the the thin S film situated at $x=0$.
The magnetic field becomes a sum of the magnetic field induced by
the domain walls and the magnetic field of the vortex. The free
energy can be written as
\begin{equation}
F = \int d^3 \mathbf{r} \left (\frac{\mathbf{B}^2}{8\pi} +
\frac{\lambda^2}{8\pi}|\nabla\times \mathbf{B}|^2 -
\mathbf{M}\mathbf{B}\right ),
\end{equation}
where $\mathbf{B}$ denotes the total magnetic field. The
difference $\mathcal{F}$ of the free energy for the state with the
vortex and the free energy for the state without vortex is given
as follows~\cite{BC}
\begin{equation}
\mathcal{F} = \frac{\phi_0}{4\pi}H_{c_1}\lambda\left [1 -
\frac{M}{M_c}\right ],\,\,\,
M_c=\frac{H_{c_1}}{4\pi}\frac{\lambda}{d_F}\mathcal{G}(\delta,\lambda,L),
\end{equation}
where $\phi_0=c h/(2 e)$, $H_{c_1}=(\phi_0/4\pi\lambda^2)
\ln\lambda/\xi$ is the lower critical field in the thin S film
without the F film and
\begin{equation}
\mathcal{G}(\delta,\lambda,L) = \left [ \frac{2 \pi
}{L}\sum\limits_{n=0}^\infty \frac{1}{\cosh\frac{\pi
q_n\delta}{2}} \frac{\delta}{(1+2q_n \lambda)^2}\right ]^{-1}.
\end{equation}
The $\mathcal{F}$ becomes negative if $M>M_c$ and vortices can
proliferate in the S film until vortex-vortex interaction stops
it or (that is more probable) the domain wall changes to Neel domain wall type to reduce 
the free energy. In the most interesting case of a single domain wall we find
\begin{equation}\label{eq:M_C}
M_c=\frac{H_{c_1}}{4\pi}\frac{\lambda}{d_F}
  \begin{cases}
    2 \lambda/\delta, & \pi\delta/4\lambda\ll 1, \\
    1 - 32 G\lambda/(\pi^2 \delta), & \pi\delta/4\lambda\gg 1,
  \end{cases}
\end{equation}
where $G\approx 0.916$ stands for the Catalan constant. (Note, that $M_c$ given by
Eq.\eqref{eq:M_C} differs from estimates of Ref.\onlinecite{Pokrovsky}).

In conclusion, domain wall effects in ferromaget-superconductor structures are investigated.
We find the distribution of current in the superconductor induced by magnetic field of Bloch
domain walls, calculate the ``lower critical'' magnetization of the ferromagnet that provides
vortices in the superconductor.

We neglected above the proximity effect in SF structure assuming that the ferromagnet and the
superconductor are weakly coupled. Below we discuss the case when S and F are strongly
coupled. Consider a SF bilayer with a perfect SF boundary. When the superconductor  and the
ferromagnet are thin enough then the bilayer can be described as a ``ferromagnetic
superconductor'' with effective parameters:\cite{Bergeret-Efetov} the superconducting gap
$\Delta_{\rm eff}$, the exchange field $E_{\rm ex}^{(\rm eff)}$... The superconductivity
survives in this system until $E_{\rm ex}^{(\rm eff)}<\Delta_{\rm eff}^{(0)}$, where
$\Delta_{\rm eff}^{(0)}$ is the gap at $E_{\rm ex}^{(\rm eff)}=0$.\cite{Bergeret-Efetov}
Domain wall structure of the ferromagnet makes the effective exchange field nonhomogeneous. We
find that if $E_{\rm ex}^{(\rm eff)}$ changes its sign on scales of the order of $\xi_0$ or
smaller then superconductivity in the bilayer survives at $\sqrt{\langle (E_{\rm ex}^{(\rm
eff)})^2\rangle}>\Delta_{\rm eff}^{(0)}$, where $\langle (E_{\rm ex}^{(\rm eff)})^2\rangle$ is
the average square of the effective exchange field over the
sample.\cite{Chtchelkatchev-Burmistrov}

We are grateful to V. Ryazanov for stimulating discussions and also thank RFBR ~Project No.
03-02-16677, 04-02-08159 and 02-02-16622, the Russian Ministry of Science, the Netherlands
Organization for Scientific Research ~NWO, CRDF, Russian Science Support foundation and State
Scientist Support foundation (Project No. 4611.2004.2).

\end{document}